\documentclass[aps,twocolumn,showpacs]{revtex4}
\usepackage{graphicx}
\usepackage[all]{xy}
\usepackage{amsmath}
\usepackage{amssymb}
\newcommand{\be}{\begin{equation}}
\newcommand{\ee}{\end{equation}}
\newcommand{\ben}{\begin{eqnarray}}
\newcommand{\een}{\end{eqnarray}}
\newcommand{\bes}{\begin{subequations}}
\newcommand{\ees}{\end{subequations}}
\newcommand{\wt}{\widetilde}

\newcommand{\bb}{\bibitem}
\begin{document}
\title{Deformed defects with applications to braneworlds}
\author{D. Bazeia and L. Losano}
\affiliation{{\small Departamento de F\'{\i}sica, Universidade Federal da Para\'{\i}ba\\
Caixa Postal 5008, 58051-970 Jo\~ao Pessoa, Para\'{\i}ba, Brazil}\\}

\begin{abstract}
In this work we investigate two distinct extensions of the deformation procedure introduced in former works on deformed defects. The first extension deals with the use of deformation functions which can assume complex values, and the second concerns the possibility of making the deformation dependent on the defect solution of the model used to implement the deformation. These two extensions bring the deformation procedure to a significantly higher standard and allow to build interesting results, as we explicitly illustrate with examples of current interest to high energy physics.
\end{abstract}


\maketitle

\section{Introduction}
\label{s1}

The study of defect structures is of great interest to high energy physics, and have been the subject of many investigations, as we see in Refs.~{\cite{r,vs,sm}}. They also play important role in condensed matter physics, as shown for instance in Refs.~{\cite{cm1,cm2}}. 

In condensed matter, an example of importance is the presence of kink-like defects in the quasi-one-dimensional organic system trans-polyacetylene \cite{ssh,js}, which may be responsible for tremendous increase in the conductivity to almost metallic level of this insulator, when charged defects are introduced by doping \cite{mh}. A general issue concerning defect structures in condensed matter is the interpretation of spontaneous symmetry breaking as the opening of a gap in the mass spectrum of the charge carriers. Thus, if we search for a mechanism to control spontaneous symmetry breaking, we also search to control the mass gap for fermionic carriers, and this is of direct interest to applications in condensed matter. For this reason, we stress that the deformation procedure set forward in Ref.~{\cite{dd}} is of direct interest to condensed matter, because it is an important mechanism to control how spontaneous symmetry breaking arises in the system. 

Another important issue concerning defect structures in high energy physics is connected with gravity with warped geometry involving a number of extra dimensions, which directly depends on the specific defect structure under consideration. For instance, if one deals with warped geometry with one, two or three extra dimensions, the interest gets to kinks, vortices or monopoles, respectively, since kinks, vortices and monopoles are defect structures that require one, two and three spatial dimensions, respectively. Works that illustrate the subject can be found for instance in Refs.~{\cite{k1,k2,k3,k4,k4',k5,k6,k7,k8,k9,e1,e2,e3,e4,e5}}. Several distinct braneworld scenarios are connected to defects which appear
in diverse dimensions. If one considers kinks generated by scalar fields, the system may be driven by the $\phi^4$ or sine-Gordon model. But we can also deal with vortices, using the Maxwell-Higgs \cite{no,b} or Chern-Simons-Higgs model \cite{jw}. Moreover, if we deal with monopoles, we may use the 't Hooft-Polyakov model \cite{t}.

The above reasonings encourage us to further investigate the deformation procedure presented in Ref.~{\cite{dd}}, trying to bring it to a higher standard, focusing mainly on applications to braneworld scenarios with a single extra dimension. To do this, in the next Sec.~II we turn attention to defect structures which appear in $(1,1)$ space-time dimensions, that is, we investigate kinks and lumps in models described by a single real scalar field. There we present two distinct extensions for deformed defects, one which considers more general deformation functions, and the other which deals with deformations that directly depend on the static solutions supported by the model to be deformed. In Sec.~III we use some of the results obtained in Sec.~II to investigate braneworld scenarios involving thick brane with internal structure. Finally, in Sec.~IV we present our comments and conclusions.

\section{The deformation procedure}

Before dealing with the deformation procedure, let us consider two distinct models described by a single real scalar field in $(1,1)$ space-time dimensions. The first model is the starting model, which is described by the Lagrange density
\be\label{sm}
{\cal L}=\frac12\partial_\mu\phi\,\partial^\mu\phi-V(\phi)
\ee
where $V(\phi)$ is the potential. Our metric is $(+,-)$ and we work with dimensionless fields and coordinates, for simplicity. The potential has at least one critical point at $\bar\phi,$ that is, $V^{\prime}(\bar\phi)=0$, for which we also set $V(\bar\phi)=0$. The equation of motion for static field is
\be\label{em1}
\frac{d^2\phi}{dx^2}=V^{\prime}(\phi ) 
\ee
where the prime stands for derivative with respect to the argument.

The equation of motion (\ref{em1}) allows writing
\be\label{em1phi}
\frac{d\phi}{dx}=\pm\sqrt{2V+c}
\ee
where $c$ is a real constant. 

Firstly, we search for static solutions $\phi=\phi(x)$ which obey the boundary conditions:
\bes
\ben
&&\phi(x\to-\infty)\to\bar\phi
\\
&&\frac{d\phi}{dx}(x\to-\infty)\to0
\een
\ees
This implies that $c$ should vanish, leading to 
\be\label{em1phi0}
\frac{d\phi}{dx}=\pm\sqrt{2V}
\ee
giving a necessary condition for the presence of finite energy solutions.

We consider another model, the deformed model, which is described by the real scalar field $\chi$
\be\label{m2}
{\cal L}_d=\frac12\partial_\mu\chi\partial^\mu\chi-{\wt V}(\chi)
\ee
Here the static solution has to obey
\be\label{em2}
\frac{d^2\chi}{dx^2}={\wt V}^{\prime}(\chi ) 
\ee
We suppose the model supports field configurations obeying boundary conditions similar to the former ones, such that
\be
\frac{d\chi}{dx}=\pm\sqrt{2{\wt V}}
\ee
where we have set to zero the integration constant, leading to a necessary condition for the presence of static solution of finite energy.

In the first work on deformed defects, one has started with a model which supports finite energy defect structures \cite{dd}. The deformation procedure relies on using real, invertible and diferenciable deformation function, $f=f(\chi),$ from which we could write the potential ${\wt V}(\chi)$ in the form
\be
{\wt V}(\chi)=\frac{V(\phi\to f(\chi))}{[df/d\chi]^2}
\ee 
In this case, we get that static solutions to Eq.~(\ref{em2}) are given by $\chi(x)=f^{-1}(\phi(x)),$ for $\phi(x)$ being static solution
to Eq.~(\ref{em1}). In the second work, we extended the procedure with the inclusion of more general deformation functions. In the present work, however, we consider two new possibilities of extending the deformation procedure used in\cite{dd}, as we show below.

\subsection{Type-1 family of deformations}

We first deal with the case $c=0,$ which is identified by Eq.~(\ref{em1phi0}). In this case, we suppose the two models (\ref{sm}) and (\ref{m2}) are connected by deformation function which is not everywhere real-defined. We name this kind of modification as the type-1 deformation. It does not modify the deformation procedure itself, but enlarge the possibilities by changing the specific characteristics of the function we need to implement the deformation. The type-1 family of deformations was initiated in the second work in Ref.~{\cite{dd}}. We illustrate this case with some interesting examples.

Consider the starting model as the $\phi^4$ model, with the potential given by
\be\label{potphi4}
V(\phi)=\frac12(1-\phi^2)^2
\ee
where we are using dimensionless fields and coordinates. It supports the kink like solutions $\phi_{\pm}(x)=\pm\tanh(x).$

We now use as deformation one of the functions $f_{\pm}(\chi)=\sqrt{1\pm\chi}.$ Although these functions are not real-defined for $\chi$ spanning the real line, the deformed model is still well-defined: the potentials are given by
\be\label{r1}
{\wt V}_{\pm}(\chi)=2\chi^2\pm 2\chi^3
\ee 
They are of the $\chi^3$ type, and the deformed models have solutions given by $\chi_\pm(x)=\pm{\rm sech}^2(x),$ which are directly obtained from the deformation procedure. The static solutions are lumps-like solutions; they are unstable, and have found recent interest as toy model to mimic properties of tachyonic excitations of non-BPS branes in string theory \cite{z}. 

Another example is given by $f(\chi)=\sqrt{1-\chi^2}.$ In this case the deformed model has the potential
\be\label{r2}
{\wt V}(\chi)=\frac12 \chi^2(1-\chi^2)
\ee
which is an inverted $\chi^4$ model, solved by $\chi(x)=\pm{\rm sech}(x),$ as we can directly obtain from the deformation procedure.

We invent other models with the help of the more general deformation functions
\be
f_{\pm}(\chi)=\mp1\pm2\chi^n
\ee
which are good for $n$ real, positive, such that $\chi^n\in{\bf R}$ for $\chi\in{\bf R}.$ Interesting examples are given by $n=1,2,3,...,$ and
$n=1/3,2/3,1/5,2/5,....$ In the general case the potentials are given by
\be\label{r3}
{\wt V}_n(\chi)=\frac2{n^2}{\chi}^2(1-{\chi}^n)^2
\ee
They have static solutions 
\be
{\chi}_{n\pm}^{\pm}(x)=\pm\Bigl[\frac12(1\pm\tanh(x))\Bigr]^{1/n}
\ee
which are also obtained from the deformation procedure above. In Fig.~[1] we plot the potentials and their respective static solutions for $n=4$ and $n=2/3,$ to illustrate how they behave in terms of the scalar field.

\begin{figure}[!ht]
\vspace{.3cm}
\includegraphics[{height=8cm,width=4cm,angle=-90}]{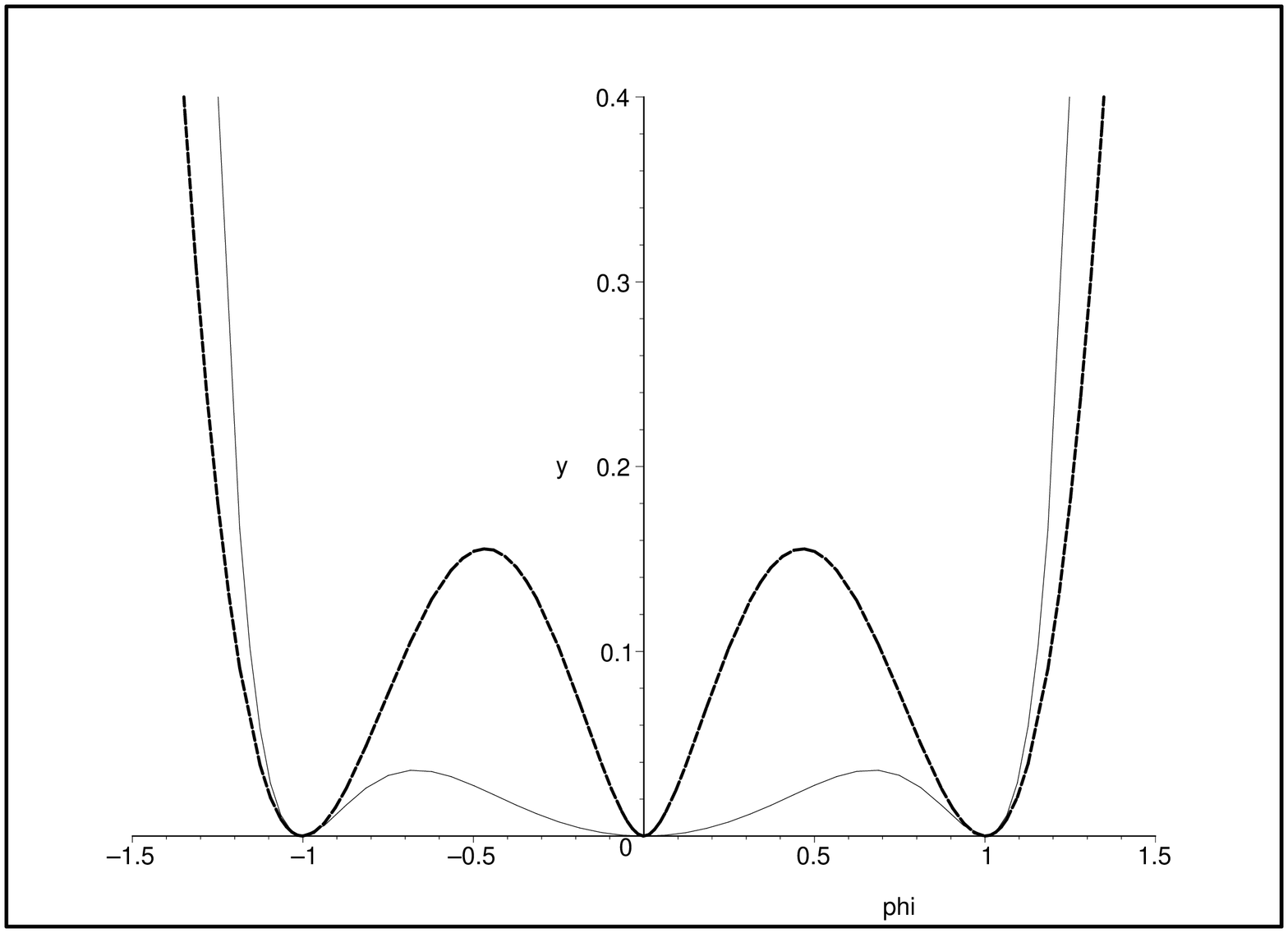}
\includegraphics[{height=8cm,width=4cm,angle=-90}]{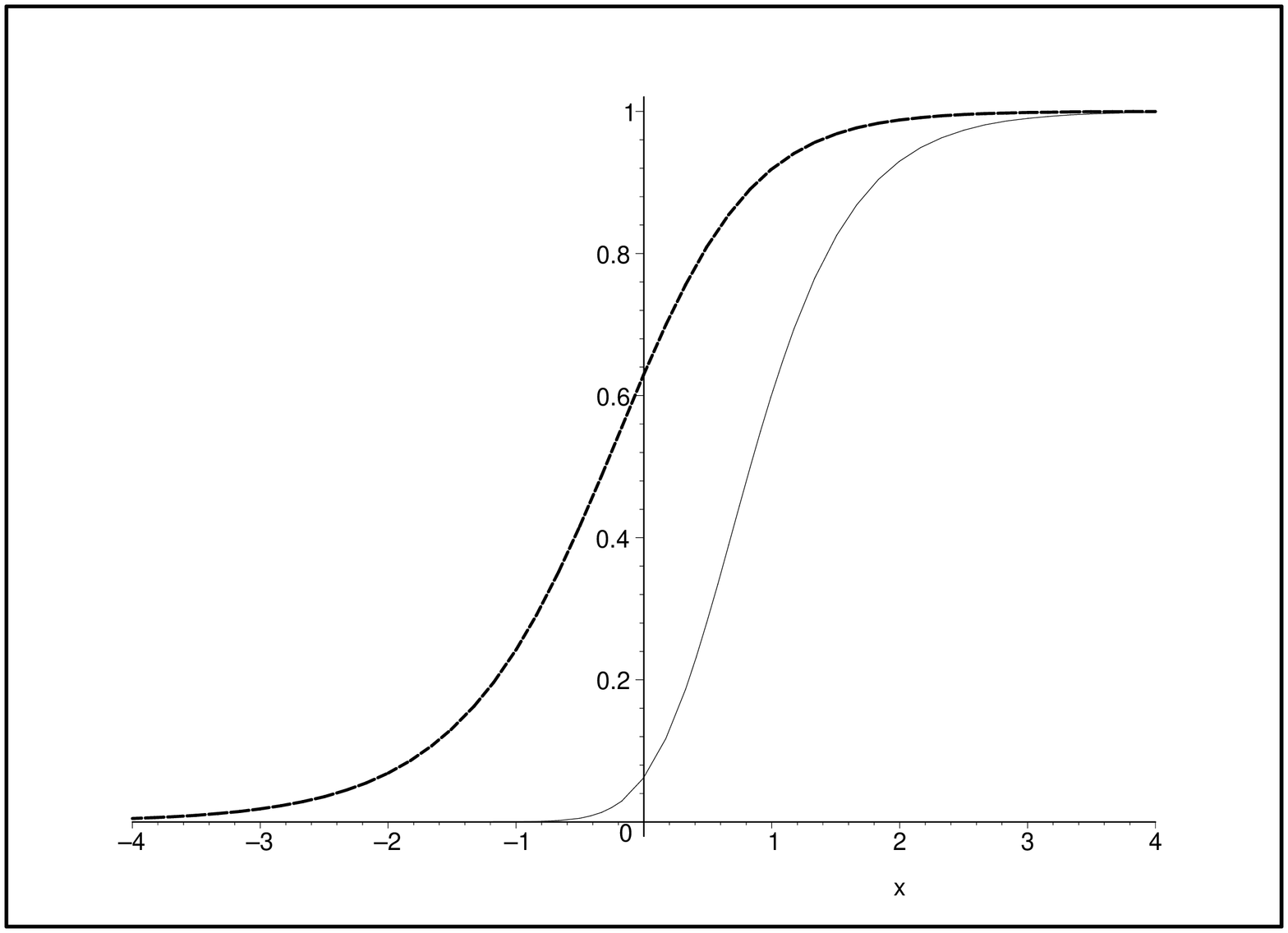}
\vspace{0.3cm}
\caption{Potentials of Eq.~(\ref{r3}) in the case of $n=4$ and $2/3$ (upper panel) and the corresponding kinks (lower panel), plotted with thinner and thicker lines, respectively.}
\end{figure}

\subsection{Type-2 family of deformations}

In the type-2 family of deformations we modify the deformation procedure itself. This is done with $c\neq0,$ starting with static solutions for which the energy diverge. This possibility was never investigated before. The system is now identified by Eq.~(\ref{em1phi}), and we can use it to get to deformed models, which support finite energy static solutions. In this case, we may start with model and static solution which are not of interest to physics, but we end up with deformed model and static solution which are well-behaved and of interest. An example of this is given by the model
\be
V(\phi)=\frac12\phi^2
\ee
This model represents the free Klein-Gordon field, and it is known to have no finite energy static solution. However, we can easily check that
it supports the set of static solutions
\be\label{staticphi}
\phi(x;a,b)=\frac12\,a\,e^{x}-\frac12\,b\,e^{-x}
\ee
where $a$ and $b$ are real parameters. These solutions should be discarded, since they are not finite energy solutions. However, we can use them to generate solutions in other models, using the deformation procedure above. We have found a nice way to get to this possibility with the prescription
\be\label{dpotchi}
{\wt V}(\chi)=\frac12\frac{2V(f)+c}{[df/d\chi]^2}
\ee
for field configurations which solve Eq.~(\ref{em1phi}). For instance, we consider the deformation function
\be\label{deff}
f(\chi)=\frac{\chi}{\sqrt{1-\chi^2}}
\ee
This gives the potential
\be\label{R1}
{\wt V}(\chi)=\frac12[ab+(1-ab)\chi^2](1-\chi^2)^2
\ee
and the static solutions
\be
\chi(x)=\pm\frac{\phi(x;a,b)}{\sqrt{1+\phi^2(x;a,b)}}
\ee
In the above manipulations, consistency between Eq.~(\ref{em1phi}) and the solution (\ref{staticphi}) imposes the restriction $c=ab,$ which makes the product $ab$ the control parameter for the potential of the deformed model. This procedure leads to three distinct scenarios. The first case is obtained for $ab\leq0,$ and we get to the class of potentials depicted in Fig.~[2a], for $ab=0,-1/2,-1$. For $ab=[0,1],$ the class of potentials goes from the $\chi^6,$ for $ab=0,$ to the $\chi^4,$ for $ab=1,$ as we show in Fig.~[2b], where we plot the potential for $ab=0,1/9,$ and $8/9.$ The third case is for $ab\geq1,$ and now the potentials are depicted in Fig.~[2c], for $ab=1,4/3,5/3.$ We illustrate the presence of static solutions in Fig.~[3], where we plot the kink-like solutions corresponding to the potentials shown in Fig.~[2b], for $ab=0,1/9,8/9.$

The potentials depicted in Fig.~[2b] were already considered in Ref.~{\cite{cl}} as examples of one-dimensional bags \cite{mit}, with the identification of quarks as defect structures. An important feature of the two-kink solution is that it does not terminate at $\phi=0,$ unless we take $ab=0,$ in this case getting to the $\phi^6$ model which supports two distinct but degenerate sectors for static solutions. Evidently, for $ab\in(0,1)$ the two-kink state approaches but is not a true two-kink, because $\phi=0$ is not a minimum of the potential. True two-kink states were recently found in Ref.~{\cite{bmm}}; they are here reproduced as the solutions show by Eq.~(\ref{2k}), for the potential of Eq.~(\ref{potp}) with $p=3,5,7,...$ A interesting issue concerns the fact that to attain the Bogomol'nyi bound, vortices in the Maxwell-Higgs model requires a $\phi^4$ potential, and in the Chern-Simons-Higgs model they need a $\phi^6$ model. Thus, we can use the above deformation to present unified picture for vortices in generalized Maxwell-Higgs model, to map the Maxwell-Higgs and Chern-Simons-Higgs models, as done in Ref.~{\cite{ba}}.  

\begin{figure}[!ht]
\vspace{.3cm}
\includegraphics[{height=8cm,width=4cm,angle=-90}]{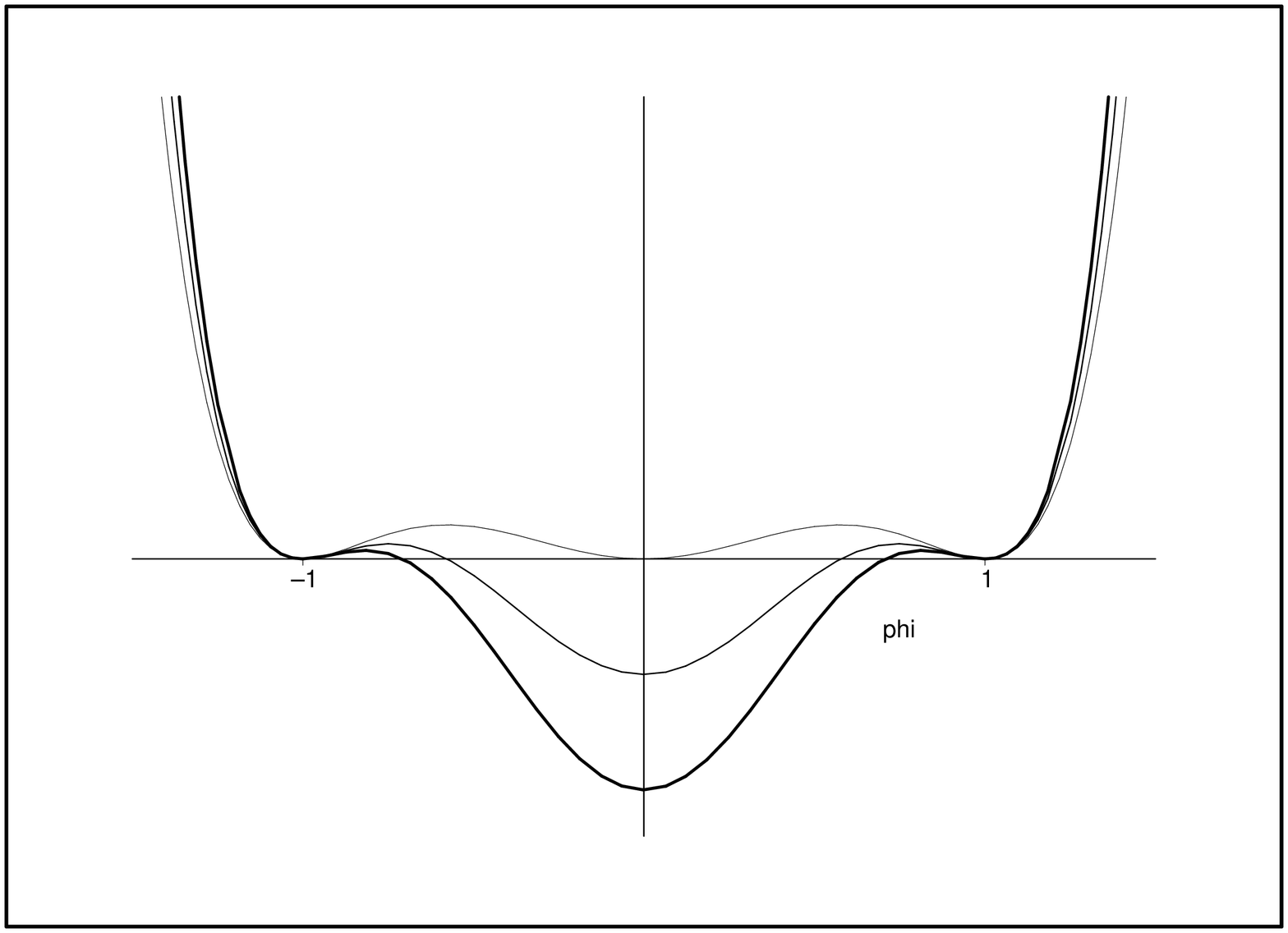}
\includegraphics[{height=8cm,width=4cm,angle=-90}]{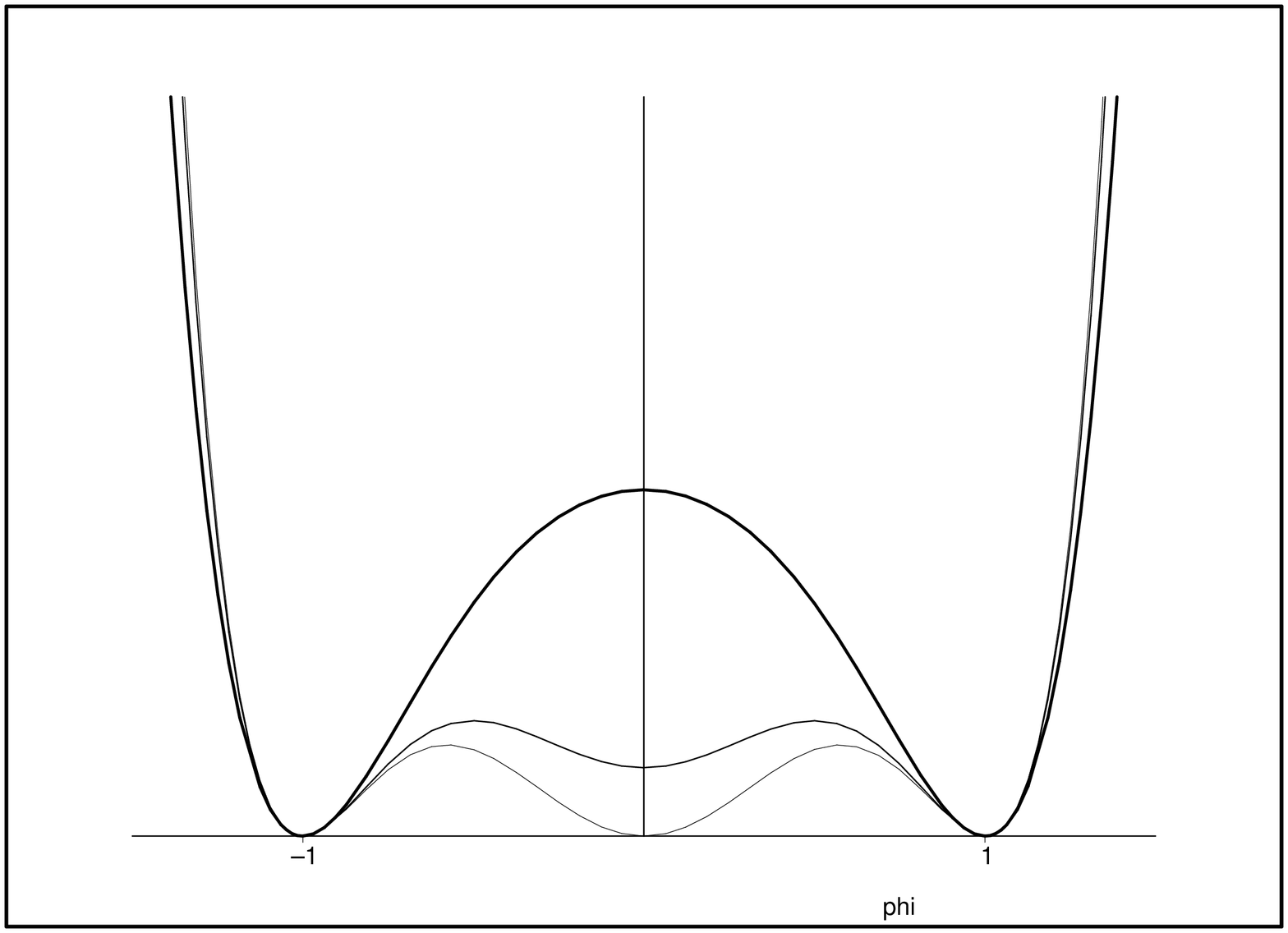}
\includegraphics[{height=8cm,width=4cm,angle=-90}]{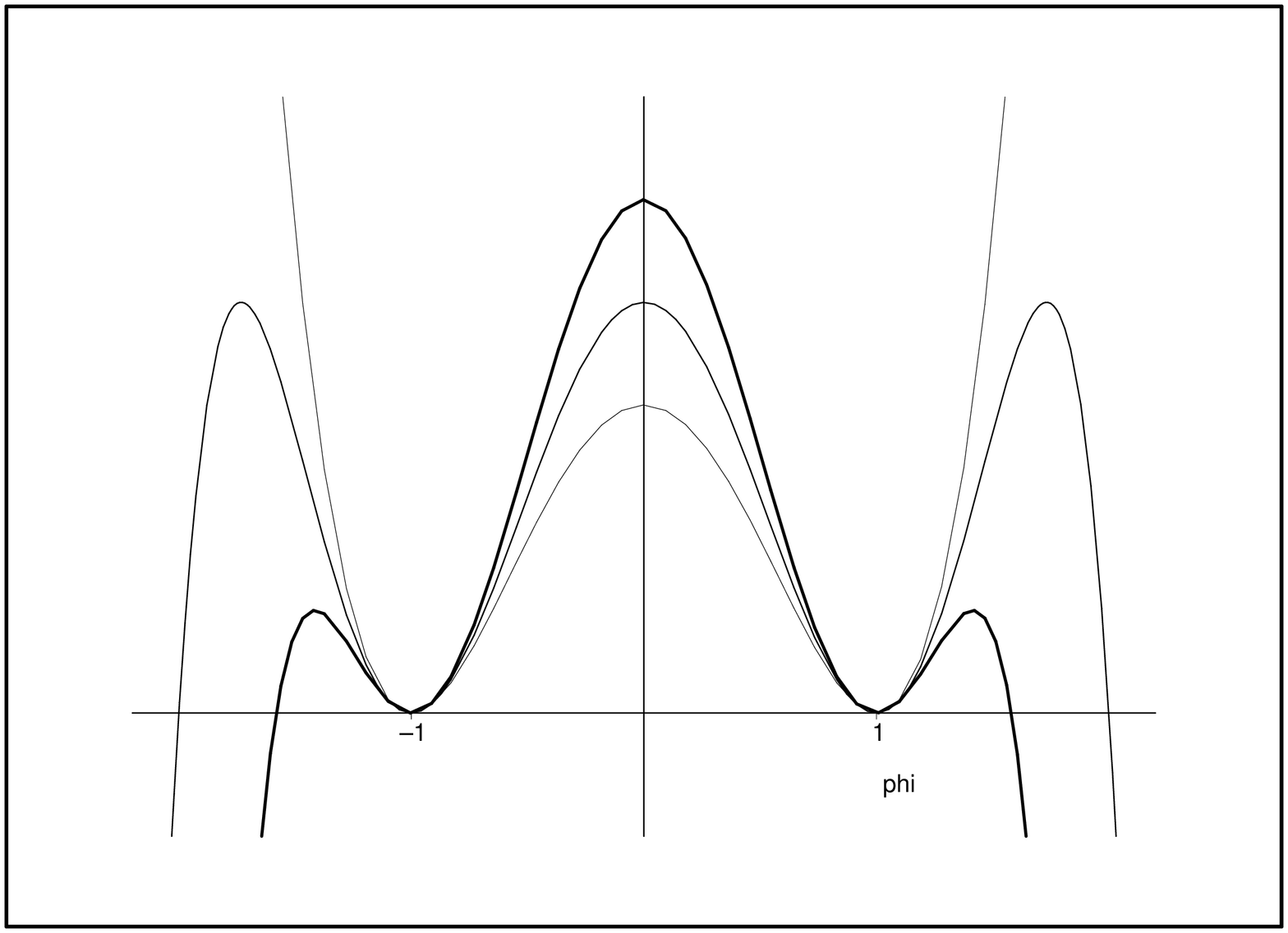}
\vspace{0.3cm}
\caption{Three distinct classes of potentials obtained as deformed potentials, depicted for $ab\leq0,$ $ab\in[0,1],$ and $ab\geq1$ in the upper, middle or lower panel, respectively. The plots correspond to the values $ab=0,-1/2,-1$, for $ab\leq0,$ $ab=0,1/9,8/9,$ for $ab\in[0,1],$ and $ab=1,4/3,5/3,$ for
$ab\geq1.$ The thickness of the lines increases with increasing $|ab|.$}
\end{figure}

\begin{figure}[ht]
\includegraphics[{height=8cm,width=4cm,angle=-90}]{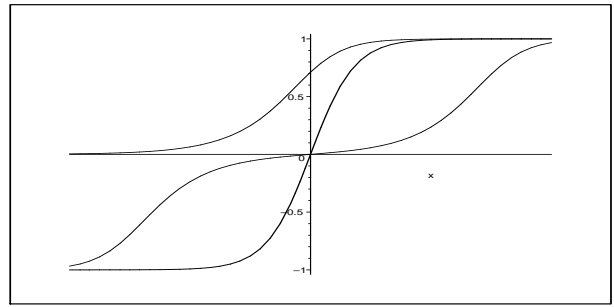}
\vspace{0.3cm}
\caption{Plots of the deformed defects corresponding to the potentials depicted in the middle panel of the former Fig.~[2]. The thickness of the lines increases with increasing $ab.$}
\end{figure}

Another class of potentials of interest can be obtained if we change the deformation function (\ref{deff}) to the new one
\be
f(\chi)=e^{{\rm arctanh}[\sin(\chi)]}
\ee
The procedure is similar to the former one, and gives the deformed potential
\be\label{defpotsg}
{\wt V}(\chi)=\frac12(ab-1)\sin^2(\chi)-ab\sin(\chi)+\frac12(ab+1)
\ee
The corresponding static solutions are obtained in the form, in the case where $a=b,$ for kink
\ben
\chi_k(x)=\left\{
\begin{array}{l}
-{\rm arcsin}[g(x)]+(2k-1)\pi,\;\;\;x\leq0
\\
{\rm arcsin}[g(x)]+2k\pi,\;\;\;x\geq0
\end{array}
\right.
\een 
and for anti-kink
\ben
\chi_{\bar k}(x)=\left\{
\begin{array}{l}
{\rm arcsin}[g(x)]+2k\pi,\;\;\;x\leq0
\\
-{\rm arcsin}[g(x)]+(2k-1)\pi,\;\;\;x\geq0
\end{array}
\right.
\een
where $k=0,\pm1,\pm2,...,$ and 
\be
g(x)=\frac{a^2\sinh^2(x)-1}{a^2\sinh^2(x)+1}
\ee
In general, the product $ab$ gives rise to three distinct classes of models, for $ab\leq0,$ $ab\in[0,1],$ and $ab\geq1.$ We illustrate these possibilities in Fig.~[4a], where we depict potentials for some values of $a$ and $b,$ such that $ab\in[0,1].$ Also, in Fig.~[4b] we plot the kinks corresponding to the potentials depicted in the upper panel. Evidently, the static solutions are all obtained via the deformation procedure given above.

As we already know, both the $\phi^4$ and sine-Gordon models can be nicely used to give rise to thick brane \cite{k4,k4',k5}. Thus, we can use the above
deformed $\phi^4,$ $\phi^6$ and sine-Gordon like models to induce modifications in the internal structure of the branes investigated in \cite{k4,k4',k5}. This study is initiated in the next section, where we show explicitly how to use deformed defects in braneworld scenarios. The subject will be part of a specific work on thick branes, in which we follow the lines of Ref.~{\cite{e2}}, taking advantage of the approach introduced in Ref.~{\cite{dd}} to investigate how to deform a model asymmetrically, to extend the investigations to the case of asymmetric branes \cite{e1,e3,e4,e5}. The motivation is interpolate between space-times with different cosmological constants, in the form of an alternative mechanism for geometric transition, well distinct from the one guided by thermal effects \cite{bbg}. 

\begin{figure}[!ht]
\vspace{.3cm}
\includegraphics[{height=8cm,width=4cm,angle=-90}]{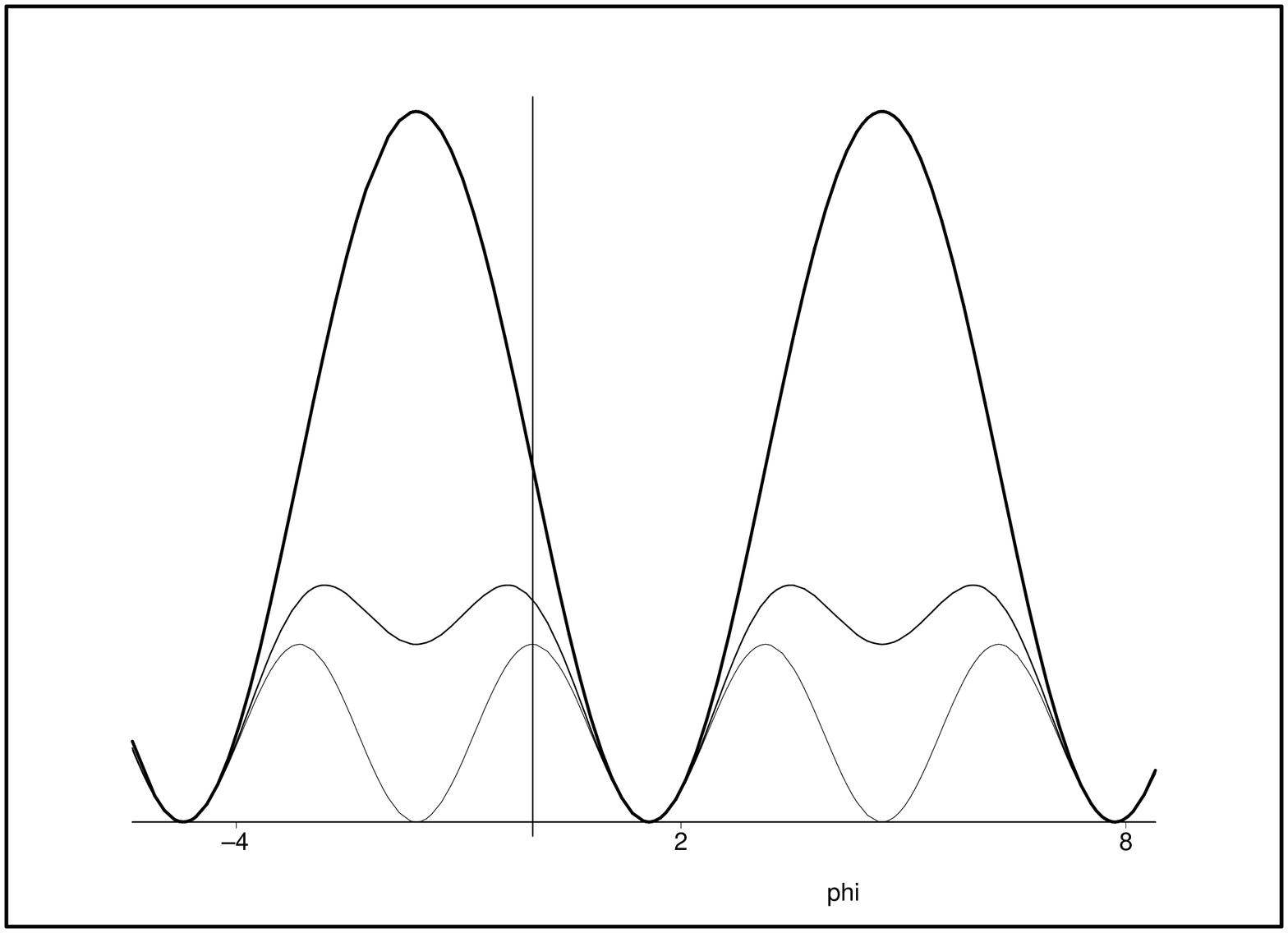}
\includegraphics[{height=8cm,width=4cm,angle=-90}]{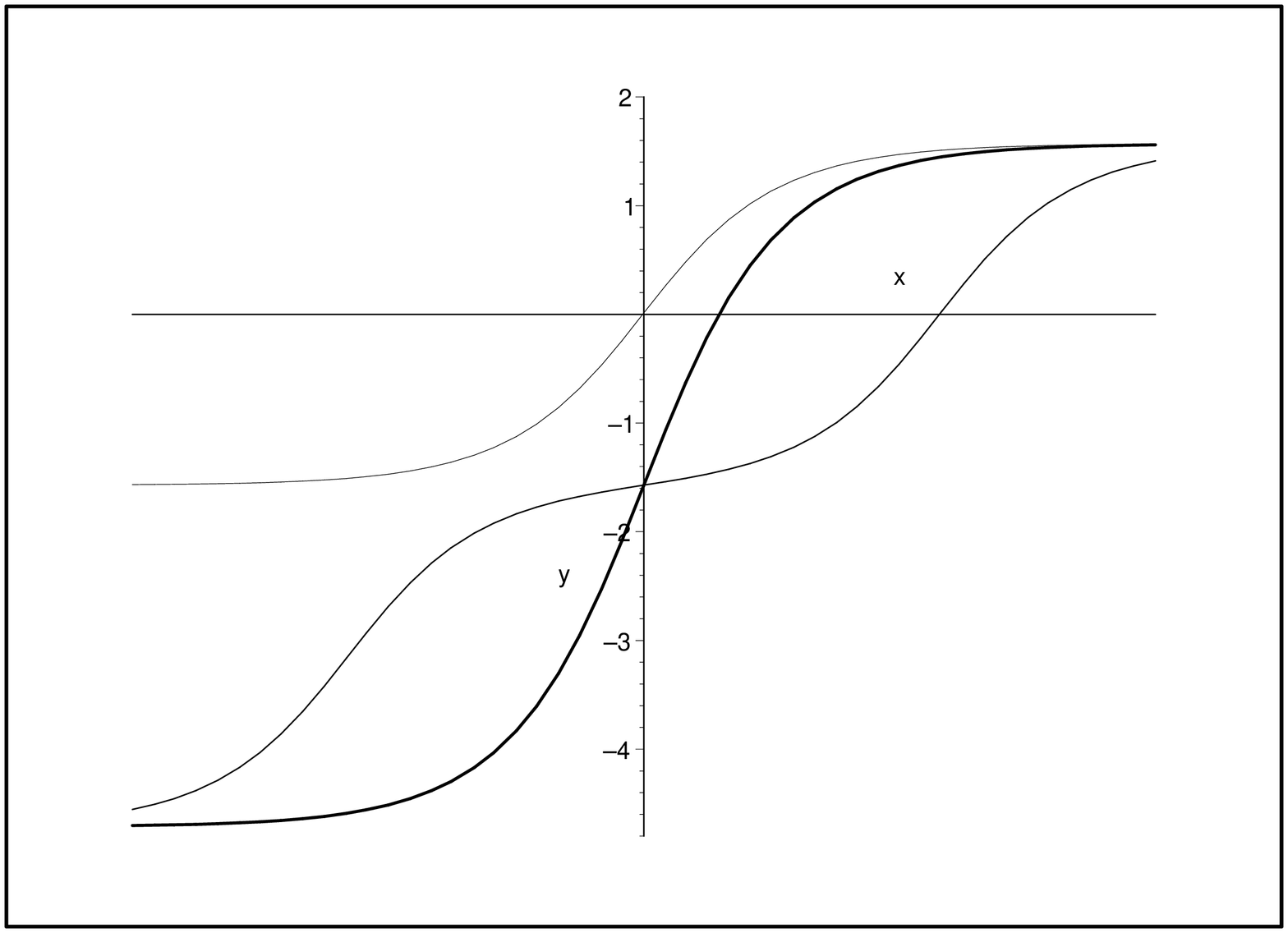}
\vspace{0.3cm}
\caption{Deformed sine-Gordon potentials obtained in Eq.~(\ref{defpotsg}), corresponding to the values $ab=0,1/4,1,$ (upper panel)
and the corresponding static kink-like solutions (lower panel). The thickness of the lines increases with increasing $ab.$}
\end{figure}

\section{Applications to braneworlds}

We now focus attention to the braneworld scenario described by a single real scalar field. In this case, the Einstein-Hilbert action for defect structures in the $AdS_5$ space-time geometry, warped with a single extra spatial dimension, can be written in the form
\be
I=\int d^4ydx \sqrt{|g|}\left({\frac14 R+{\cal L}(\phi,\partial_\mu\phi)}\right)
\ee
where $R$ is the scalar curvature, $g$ is the determinant of the metric tensor, described by the line-element
\be\label{le}
ds^2=e^{-2 A(x)}\eta_{\mu\nu}dy^\mu dy^\nu-dx^2
\ee
where $\mu,\nu$ vary from $0$ to $3,$ $\eta_{\mu\nu}$ stands for the Minkowski metric tensor, and $\exp(-2A)$ is the warp factor. Here $x$ represents
the extra dimension.

\begin{figure}[!ht]
\vspace{.3cm}
\includegraphics[{height=8cm,width=4cm,angle=-90}]{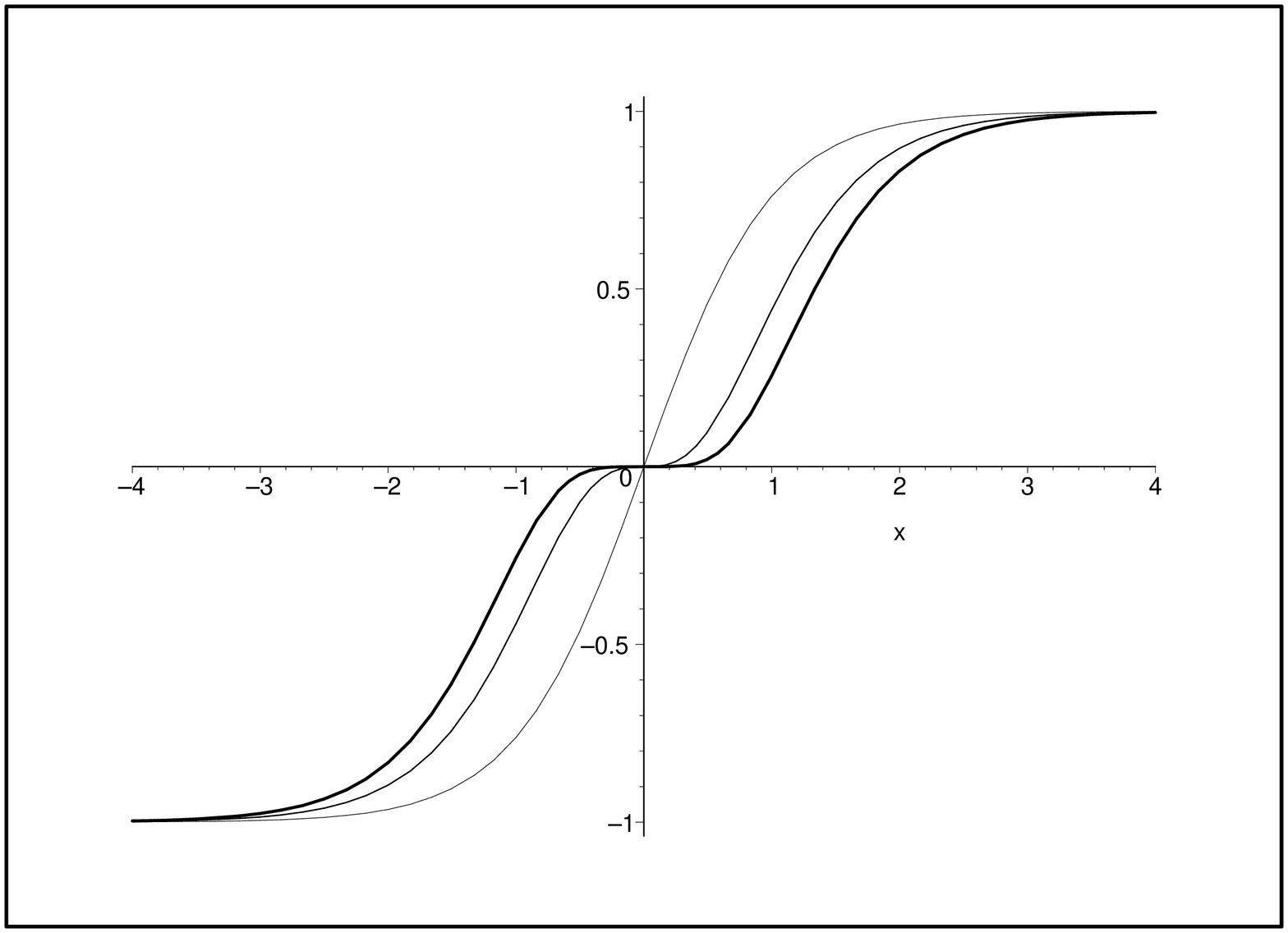}
\includegraphics[{height=8cm,width=4cm,angle=-90}]{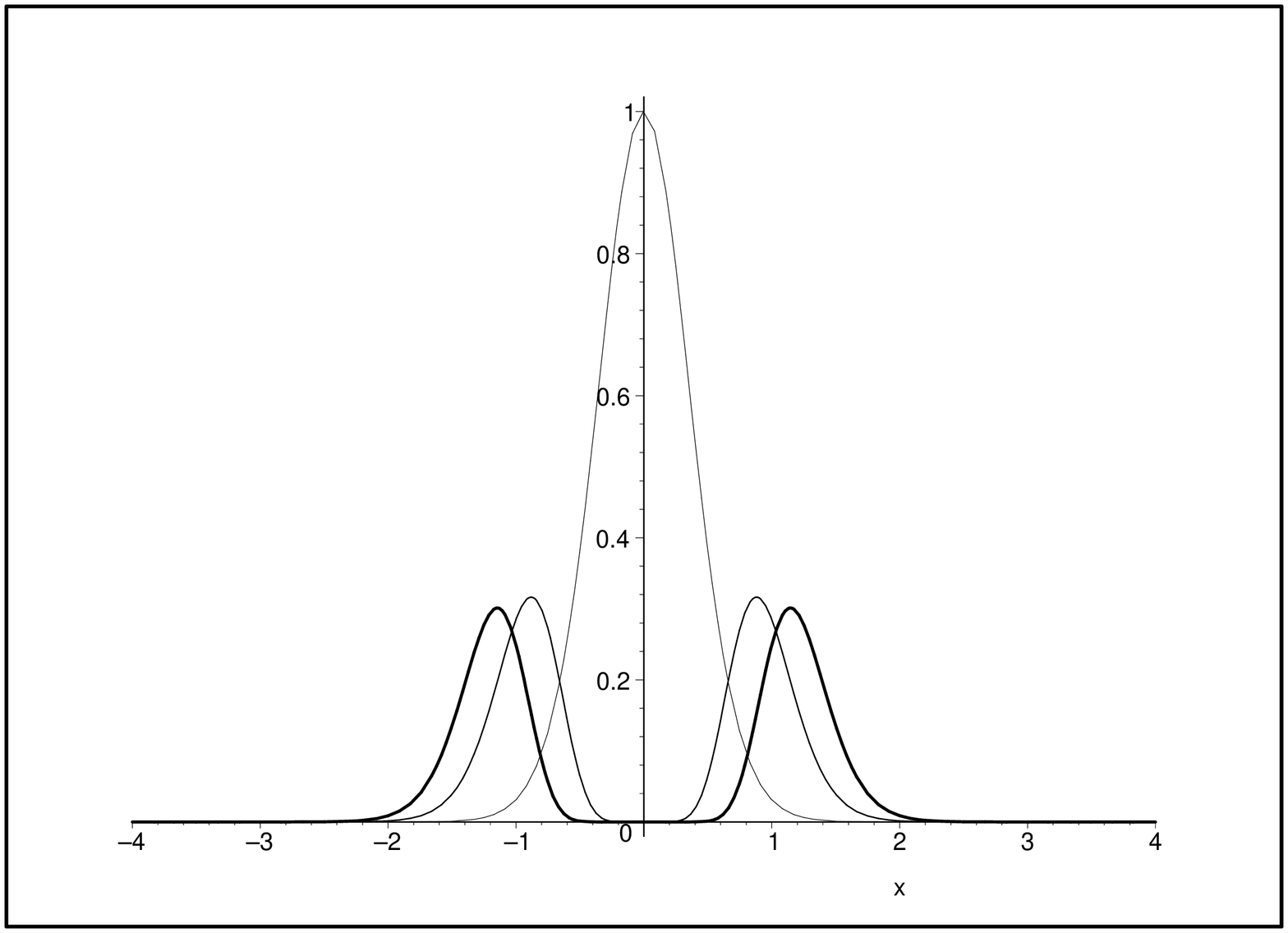}
\vspace{0.3cm}
\caption{Kink solutions of Eq.~(\ref{2k}) for $p=1,3,5$ (upper panel) and the corresponding energy densities (lower panel) in flat space-time. The thickness of the lines increases with increasing $p.$}
\end{figure}

For scalar fields, an interesting work \cite{k6} has recently shown how to generate thick brane with internal structure dependent on the temperature.
This thick brane behavior has inspired another work \cite{e2}, in which one shows how to mimic the temperature effects with unusual self-interactions, described by the potential
\be\label{potp}
V_p(\phi)=\frac12(\phi^{(p-1)/p}-\phi^{(p+1)/p})^2
\ee
invented in Ref.~{\cite{bmm}}. An important point here is that the self-interactions has been built with the help of the deformation procedure of Ref.~{\cite{dd}}. The issue is that we can deform the $\phi^4$ potential of Eq.~(\ref{potphi4}) with the deformation
\be
f(\chi)=\chi^{\frac1p}
\ee 
which behaves appropriately for $p$ odd integer, $p=1,3,5,...,$ leading to the deformed model, driven by the potential
\be
V(\chi)=\frac12p^2(\chi^{(p-1)/p}-\chi^{(p+1)/p})^2
\ee
which presents the kink-like solutions
\be\label{2k}
\chi(x)=\tanh^p(x)
\ee
These solutions are plotted in Fig.~[5a] for $p=1,3,5.$ Also, in Fig.~[5b] we depict the energy densities of the solutions given in Fig.~[5a]. We notice that $p=3,5,...$ represents two-kink solutions, and the corresponding energy densities show the appearance of internal structure, which persists
in the braneworld scenario investigated in Ref.~{\cite{e2}}.

We now consider the deformation (\ref{R1}), for $ab\in[0,1].$ This modifies the $\phi^4$ model as shown in Fig.~[2b]. We use these models in the braneworld scenario with warped geometry involving an extra, infinity dimension. We illustrate the calculation with the case $a=b,$ for $a^2=1,1/2,$ and $1/4.$ 

\begin{figure}[!ht]
\vspace{.3cm}
\includegraphics[{height=8cm,width=4cm,angle=-90}]{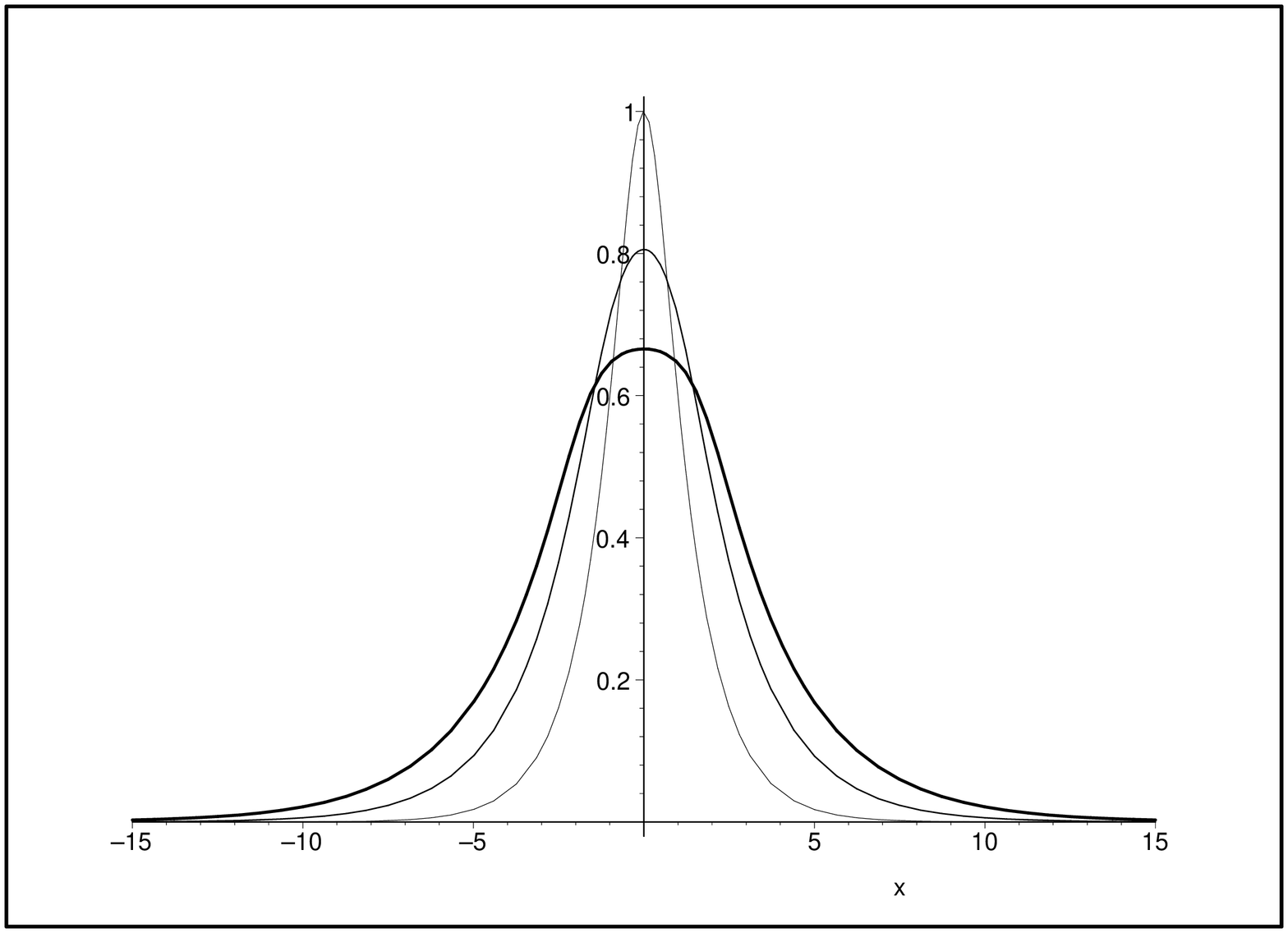}
\includegraphics[{height=8cm,width=4cm,angle=-90}]{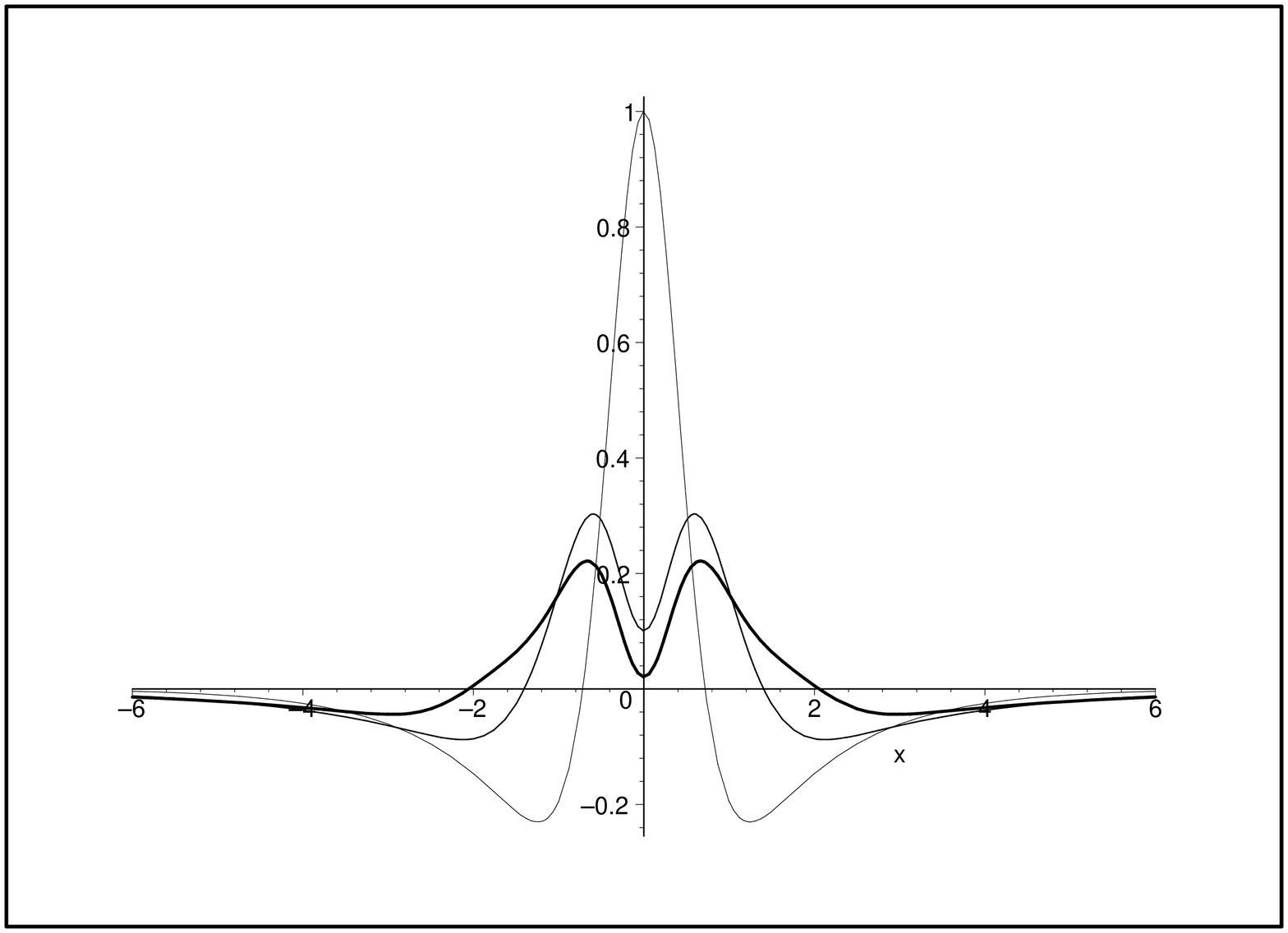}
\vspace{0.3cm}
\caption{Warp factor for $a^2=1,1/2,$ and $1/4$ (upper panel) and the corresponding energy densities (lower panel) for the scalar field in curved space-time for the model described by Eq.~(\ref{R1}) with $a=b$. The thickness of the lines increases with decreasing $a.$}
\end{figure}

In the presence of gravity, we modify the potential according to \cite{k4}
\be
V(\chi)=\frac18 \left(\frac{dW}{d\chi}\right)^2-\frac13 W^2
\ee
where $dW/d\chi=2(1-\chi^2)\sqrt{1+a^2\chi^2}\,.$ This form of potential allows the presence of first-order equations, which are given by
\be
\chi^\prime(x)=\frac12 \frac{dW}{d\chi},\;\;\;\;\;\;\;A^\prime(x)=-\frac13 W
\ee
We notice that gravity does not modify the static field $\phi(x),$ but it gives rise to warp factor which we plot in Fig.~[6] together with the corresponding energy density for the scalar field in curved space-time. There we see the warp factor getting thicker, and the energy density splitting into two distinct parts, showing the opening of a gap inside the brane. Here the mechanism is similar to the one presented in \cite{e2}, but it is different from the case explored in Ref.~{\cite{k6}}, which relies on the presence of thermal effects. 

\section{Ending comments}

In this work, we have brought the deformation procedure introduced in Ref.~{\cite{dd}} to a higher standard, including two distinct extensions, the type-1 and type-2 family of deformations. These extensions rely on modifications of the deformation function in the type-1 case, and of the deformation procedure itself for the type-2 family. The investigations were motivated by the possibilities of applications both in condensed matter and in high energy physics. In condensed matter, the deformation procedure may provide a new way to control spontaneous symmetry breaking, giving rises to a mechanism which can be used to tune the mass gap of the fermionic charge carriers. In high energy physics, an important motivation springs concerning braneworld scenarios produced by coupling gravity with scalar fields, in the warped $AdS_5$ geometry with a single extra spatial dimension. The applications presented in Sec.~III show how the deformation parameters induce modifications in the internal structure to the brane.

The present investigations have shown several results, in which one obtains interesting generalizations given by the potentials (\ref{r1}), (\ref{r2}), and (\ref{r3}), together with their respective defect structures. The type-2 family of deformations also gives important results, as we show with the generalizations obtained with the potentials (\ref{R1}) and (\ref{defpotsg}), together with their respective defect structures. These deformations can certainly be used in several contexts, for instance in the recent investigations concerning the presence of tachyons on branes from $\phi^4$ and Abelian Higgs sphalerons, which appear when one compactifies extra dimensions \cite{br,bh}, and in the general braneworld scenario with a single extra dimension, as explicitly illustrated in Sec.~III.  

We would like to thank L.A. Ferreira and J.M.C. Malbouisson for comments and discussions. We also thank CNPq, FAPESQ/CNPq/MCT, and PADCT/CNPq/MCT for financial support.


\end{document}